\documentclass[letterpaper]{JHEP3}

\usepackage{graphicx}
\usepackage{afterpage}
\def\simgt{\mathrel{\lower2.5pt\vbox{\lineskip=0pt\baselineskip=0pt
           \hbox{$>$}\hbox{$\sim$}}}}
\def\simlt{\mathrel{\lower2.5pt\vbox{\lineskip=0pt\baselineskip=0pt
           \hbox{$<$}\hbox{$\sim$}}}}

\title{Heavy Squarks at the LHC}
\author{}

\author{JiJi Fan$^a$, David Krohn$^b$, Pablo Mosteiro$^a$, Arun M. Thalapillil$^c$, and Lian-Tao Wang$^{a,c,d}$\\
        $^a$Department of Physics, Princeton University, Princeton, NJ, 08540\\
        $^b$Department of Physics, Harvard University, Cambridge MA, 02138 \\
        $^c$Enrico Fermi Institute and Department of Physics, University of Chicago, Chicago, IL, 60637\\
        $^d$Kavli Institute for Cosmological Physics, University of Chicago, Chicago, IL, 60637\\
       E-mail: \email{jijifan@princeton.edu,dkrohn@physics.harvard.edu,\\pablo@princeton.edu,madhav@uchicago.edu,liantaow@uchicago.edu}}

\abstract{
The LHC, with its seven-fold increase in energy over the Tevatron, is capable of probing  regions of SUSY parameter space exhibiting qualitatively new collider phenomenology.  
Here we investigate one such region in which first generation squarks 
are very heavy compared to the other superpartners. We find that  the production of these squarks, which is dominantly associative, only becomes rate-limited at $m_{\tilde{q}}\gtrsim 4(5)~{\rm TeV}$ for ${\cal L}\sim 10 (100)~{\rm fb}^{-1}$.  However,  discovery of this scenario is complicated because 
heavy squarks decay primarily into a jet and boosted gluino, yielding a dijet-like topology with missing energy (MET) pointing along the direction of the second hardest jet.  The result is that many signal events are
 removed by standard jet/MET anti-alignment cuts designed to guard against jet mismeasurement errors.  We suggest 
replacing these anti-alignment cuts with a measurement of jet substructure that can significantly extend the reach of this channel while still removing much of the background.  We study a selection of 
 benchmark points in detail, demonstrating that $m_{\tilde{q}}=4(5)~{\rm TeV}$ first generation squarks can be discovered at the LHC with  ${\cal L}\sim 10(100){\rm fb}^{-1}$.}

\newcommand{\beq}{\begin{equation}}% can be used as {equation} or {eqnarray}
\newcommand{\eeq}{\end{equation}}
\newcommand{\beqs}{\begin{eqnarray}}% can be used as {equation} or {eqnarray}
\newcommand{\eeqs}{\end{eqnarray}}

\newcommand{\Eq}[1]{Eq.~(\ref{#1})}
\newcommand{\Sec}[1]{Sec.~\ref{#1}}

\newcommand{\Fig}[1]{Fig.~\ref{#1}}
\newcommand{\Tab}[1]{Table~\ref{#1}}
\newcommand{\Ref}[1]{Ref.~\cite{#1}}
\newcommand{\Refs}[1]{Refs.~\cite{#1}}
\newcommand{\met}{E\!\!\!/_T}

\newcommand{\sq}{\tilde{q}}
\newcommand{\gl}{\tilde{g}}

\begin{document}
\section{Introduction}
\label{sec: introduction}
The Large Hadron Collider (LHC) has commenced operation and is already producing collisions at energies far above the scale of electroweak symmetry breaking (EWSB).  
Thus one can be hopeful that soon the particle responsible for EWSB (e.g. the Higgs) will be discovered, as well as any physics beyond the standard model (BSM) responsible for 
setting the scale at which this breaking occurs.

Supersymmetry (SUSY) is perhaps the most promising candidate for BSM physics, and as such its phenomenology has been studied extensively.  However, the SUSY parameter space is so complicated that most collider studies confine themselves to studying a restrictive subset of models or simple benchmark points (e.g. mSURGRA\cite{Chamseddine:1982jx} and the SPS points~\cite{Allanach:2002nj}, respectively).  In general though, it is useful to look at other regions of parameter space to make sure no signals are missed~\cite{Berger:2008cq,Conley:2010du}.  

Here we will focus on one such understudied region in which first generation squarks are very heavy compared to the other SUSY superpartners\footnote{
We focus on first generation squarks because they result mostly from the scattering of valence quarks, and are thus produced in much greater numbers than squarks of the second and third generation. }.
This scenario deserves attention not only because it yields a viable SUSY spectrum, but also because it is motivated 
by many interesting SUSY scenarios~\cite{Barbieri:1995uv,Cohen:1996vb} including more recent work in split SUSY~\cite{ArkaniHamed:2004fb, Giudice:2004tc, ArkaniHamed:2004yi}, PeV-scale SUSY~\cite{Wells:2004di}, and single sector SUSY breaking~\cite{Franco:2009wf, Craig:2009hf}. Intuitively, it is reasonable to anticipate heavy first generation squarks because
they play a minimal role in solving the hierarchy problem (i.e. stabilizing the electroweak scale) - only the third generation really needs to be light if SUSY plays the role we expect of it.
  In any case, here our main goal is to determine an optimal search strategy for discovering heavy squarks and to assess the reach of the LHC in finding them.

Now, when squarks are very heavy it becomes too costly to create them in pairs and so heavy squarks are dominantly produced in association with a different superpartner.  This associated particle is usually the gluino, due to its large color charge.  Furthermore, once produced the squark usually decays into a gluino and a jet, again because of the gluino's large color charge, yielding the topology:
\beqs
pp\to &\sq \gl \\
&\hookrightarrow&2\gl + j. \nonumber  \nonumber 
\eeqs
Since, in most plausible SUSY spectra, all superpartners decays yield a stable neutral particle (i.e. the LSP, labeled $\tilde{\chi}_1^0$) squark associated production would seem to yield the classic new physics signal: jets +$\met$. However, for heavy squark scenarios, new challenges appear in this well-studied final state. 

Difficulties arise because the gluino from the decaying squark will be be very energetic (assuming $m_{\tilde{g}} \ll m_{\tilde{q}}$),  and so all of its decay products become collimated, confined to a cone of opening angle
\beq
\Delta R \sim \frac{2 m_{\tilde g}}{p_T}\sim \frac{m_{\tilde g}}{m_{\tilde q}}.
\eeq
In practice, this means that the boosted  gluino's decay products will often be resolved as a single jet (henceforth the {\it gluino jet}).  Furthermore, 
although there are two sources of $\met$ in this process, the $\tilde{\chi}_1^0$ from the boosted gluino is much harder than that of the associated gluino, and 
so the $\met$ in this process tends to be aligned with the gluino jet.  Thus, what is really a complicated SUSY process is resolved as two back-to-back jets with $\met$ aligned along one of them - the telltale signature of a mismeasured QCD dijet event.
Indeed, precisely because our signal events look so much like QCD dijets they are often vetoed by standard pre-selection cuts.
Traditionally one requires a  separation in azimuthal angles between the $\met$ and any nearby jets in order to remove mismeasurement backgrounds: D0 requires $\Delta_\phi(\met, j) > 0.8$ in its  jets +  $\met$ search~\cite{:2007ww}, while the LHC experiments require $\Delta_\phi(\met, j) \gtrsim 0.3$~\cite{Ball:2007zza}.  We therefore expect to run into trouble using standard analyses when $m_{\tilde{g}}/m_{\tilde{q}} \lesssim 0.3$.

It is clear that to resolve heavy squarks at the LHC the cut on $\Delta\phi(\met, j)$ should be relaxed and some other tool must be used to remove mismeasured QCD.  Here we propose using jet substructure\footnote{For a review of these techniques, see  \Refs{Abdesselam:2010pt,Salam:2009jx}.} for this purpose.  This approach is motivated by the fact that while the four-momenta of a gluino jet could be similar to that 
of a QCD jet, the distribution of the constituent four-momenta {\it inside} the jet (i.e. the calorimeter cells inside of it) will be very different.  Gluino jets contain many hard, widely seperated subclusters of energy, evidence of a decay chain, while the depositions within a QCD jet assume a hierarchical structure, evidence for emissions governed by the showering of partons within QCD.   The variables of jet substructure allow one to distinguish between these two cases, as we will soon see.

The paper is structured as follows. In Sec.~\ref{sec: general} we discuss squark production rates and decay channels. Sec.~\ref{sec: analysis} studies the basic kinematic cuts necessary to discover heavy squarks over the SM backgrounds and discusses jet substructure observables helpful in reducing mismeasurement errors. Finally, we conclude in Sec.~\ref{sec: conclusion}.

\section {Production Rates and Branching Ratios}
\label{sec: general}
In determining the reach of the LHC in probing heavy squarks the place to start, of course, is with production rates.  As mentioned in the introduction, heavy
squarks are dominantly produced in association with one of the lighter superpartners simply because pair production would require a very high energy, the probability
of which is suppressed by the fast-falling nature of the proton's parton distribution functions (PDFs).   Furthermore the state produced in association with the squark is almost always a gluino because of its large coupling.

In \Fig{fig: SD constraint} we plot the leading order\footnote{The NLO calculation is given in~\cite{Beenakker:1996ch}.} cross section for the process $pp\rightarrow \tilde{q}+\tilde{g}$ over a range of $(m_{\tilde{q}},m_{\tilde{g}})$ and at 
different LHC energies.  These results are obtained using \texttt{Pythia v6.423}~\cite{Sjostrand:2006za} assuming the first generation squarks are all degenerate in mass and any mixings are negligible. If we require $S/\sqrt{B}> 5$ with at least 10 signal events~\cite{atlastdr2} for discovery, then one can see that for ${\cal L}\sim 10(100)~{\rm fb}^{-1}$ a 14 TeV LHC is in principle capable of discovering squarks up to $m_{\tilde{q}}\sim 4(5)~{\rm TeV}$.

\FIGURE[p!]{
 \includegraphics[scale=.6]{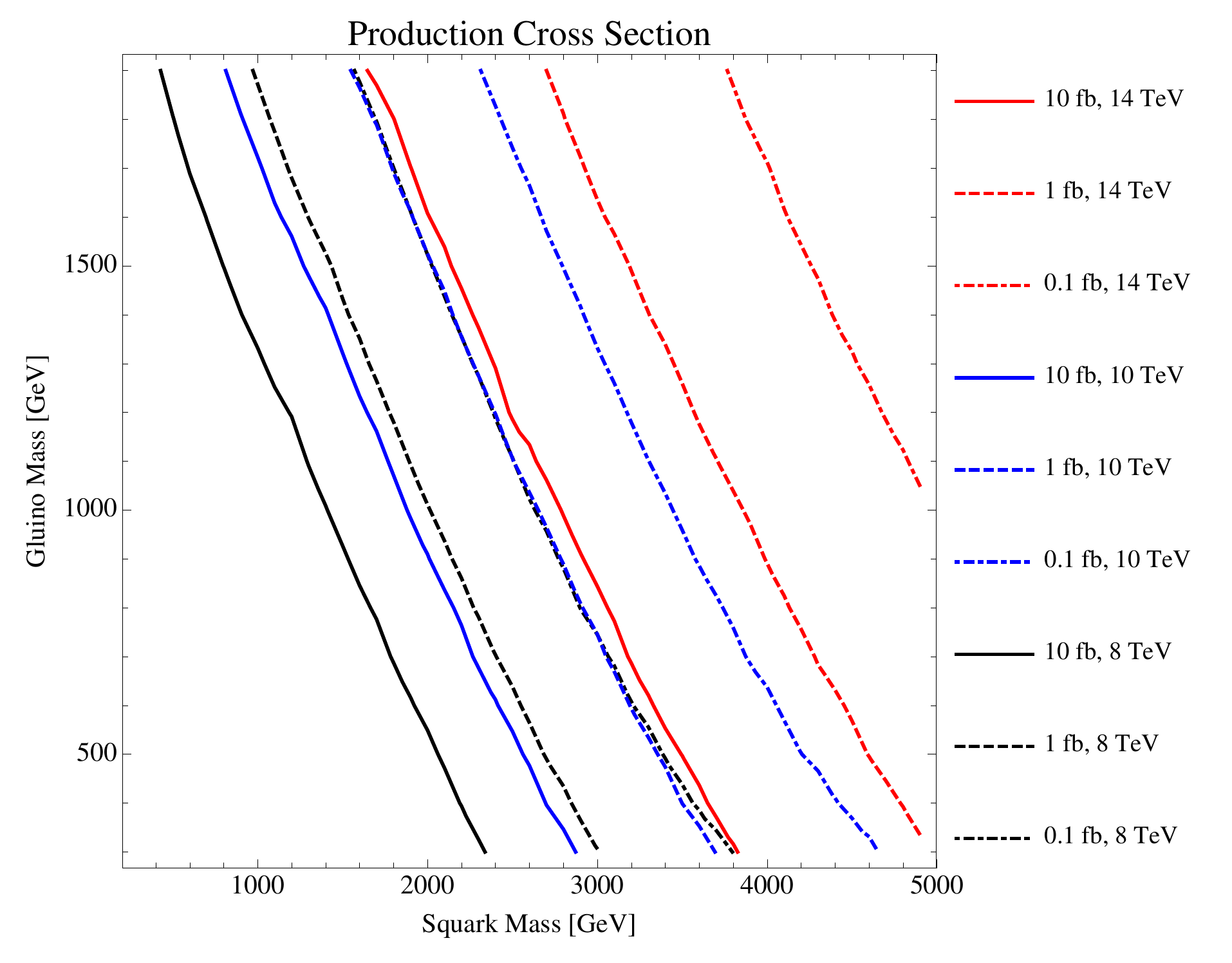}
\caption{The leading order cross-section for $\tilde{q}+\tilde{g}$ associated production at different $(m_{\tilde{q}},m_{\tilde{g}})$ for a $\hat{s}=14,10,8~{\rm TeV}$ LHC 
(red, blue, black) where $\sigma=10,1,0.1~{\rm fb}$ boundaries are indicated by solid, dashed, and dot-dashed lines, respectively.}
\label{fig: SD constraint}}

Once heavy squarks are produced they will dominantly decay through the QCD channel $\tilde{q} \to q \tilde{g}$ which will be the focus of our study, although some squarks will decay into  neutralinos or charginos\footnote{In our simulations we will account for the suppressed branching $\tilde{q}\rightarrow\tilde{g}+q$ due 
to decays into electoweak superpartners.  However, the effects of this suppression are small.}.  
Of course, once gluinos have been produced they too decay.  Fortunately, gluino decays in models with heavy squarks have been well studied in the literature~\cite{Toharia:2005gm,Gambino:2005eh}. In general, there are three major decay chains for the gluino: 
\begin{enumerate}
\item {Via an off-shell squark to two quarks and the  LSP: $\tilde{g} \to q \bar{q}+\tilde{\chi}_1^0$.  }
\item{Via an off-shell squark to two quarks and a chargino or heavier neutralino: e.g. $\tilde{g} \to q \bar{q}+\tilde{\chi}_1^+ $.}
\item{Through a squark-quark loop: $\tilde{g} \to g+ \tilde{\chi}_1^0$.}
\end{enumerate}
The explicit formulae of the decay widths can be found in \Ref{Toharia:2005gm}. A few remarks are in order: 
\begin{itemize}
\item We note that the $q\bar q$ pair formed from the decay of the gluino can,  if kinematically allowed,   be composed of top quarks.  This results in
an even richer jet substructure than the decay to light quarks.  In fact, as one expects the stop squark to be relatively light (as it plays a large role in stabilizing the Higgs mass),
and the decay widths scale as $1/m_{\tilde{q}}^4$, the branching ratio to the ditop channel $\tilde{g} \to t \bar{t}+\tilde{\chi}_1^0$ can often dominate.
\item{In the limit where $\tilde{\chi}_1^0$ is purely a Bino,  the squark mass matrices are flavor-diagonal, and left-right mixings are negligible, the branching ratio between the 2-body and 3-body ratios $R_{2/3} \equiv \Gamma(\tilde{g} \to$ 2 body)/ $\Gamma(\tilde{g} \to$ 3 body) is 
\beq
R_{2/3} \propto \sum_{flavors}\frac{\alpha_s(m_{\tilde{L}}^2-m_{\tilde{R}}^2)^2}{m_{\tilde{L}}^4+m_{\tilde{R}}^4}
\eeq
In this limit, the loop-induced decay $\tilde{g} \to g+ \tilde{\chi}_1^0$ is suppressed by the mass splitting between the left- and right- handed squarks $m_{\tilde{L}}^2-m_{\tilde{R}}^2$. This is because $\tilde{g} \to g + \tilde{B}$, governed by the magnetic operator $\bar{\chi}_1^0 \sigma^{\mu\nu} \gamma_5 \tilde{g} G_{\mu\nu}$, is a C-violating process. It vanishes for pure strong interactions with degenerate squark masses as these preserve C-parity. Any such C-violating decay channel can only be generated by the weak interactions which generate the mass splittings between squarks of different handedness.}

\item{Finally, another  interesting limit emphasized in~\cite{Toharia:2005gm} is when the gluino is kinematically allowed to decay into Higgsinos. Here the radiative decay is enhanced by $(log(m_{\tilde{t}}/m_t))^2$ due to the stop and top loop.} 
\end{itemize}
For our study, we will take the mass differences between $m_{\tilde{L}}^2-m_{\tilde{R}}^2$ to be negligible compared to the squark mass, and we will posit that gluino decays to Higgsinos are kinematically forbidden.  Thus, in what follows we will be considering only gluino three-body decays.

\section{Analysis}
\label{sec: analysis}
To study the phenomenology of the different gluino decay possibilities discussed in \Sec {sec: general} we choose the three benchmark models presented in \Tab{table:models}.  We will study each of these points at $m_{\tilde{q}}=4~{\rm TeV}$ and  $5~{\rm TeV}$.
\begin{table}
\begin{center}
\begin{tabular}{|c|c|c|c|c|}
\hline 
Model&$m_{\tilde{g}}$ [GeV] & $m_{\tilde{\chi}_1^0}$ [GeV] & ${\tilde{g}}$ decay channel & Br($\tilde{q}\rightarrow\tilde{g}+q$)[\%] \\
\hline
1&400 & 150 &$\tilde{g} \to jj \tilde{\chi}_1^0$& 88 (88)\\
\hline
2&400 &100 &  $\tilde{g} \to jj \tilde{\chi}_{2,3,4}^0 \to jj\tilde{\chi}_1^0 Z/h$ & 78 (77)\\
&&& $\tilde{g} \to jj \tilde{\chi}_{1,2}^\pm \to jj\tilde{\chi}_1^0 W^\pm$&\\
\hline
3&600 & 150 &$\tilde{g} \to t\bar{t} \tilde{\chi}_1^0$ & 76 (82)\\
\hline
\end{tabular}
\caption{The three benchmark SUSY models we will consider.  In the last column, 
where we have listed the branching ratio for the decay of the squark into a gluino,
the number outside of parenthesis is for $m_{\tilde{q}}=4~{\rm TeV}$ and the number
inside is for $m_{\tilde{q}}=5~{\rm TeV}$.}
\label{table:models}
\end{center}
\end{table}
These points include the decay of the gluino into two light jets+$\tilde{\chi}_1^0$, the cascade decay into heavy electroweak particles along with two jets and a $\tilde{\chi}_1^0$, and the 
decay into top pairs with a $\tilde{\chi}_1^0$.
All of these points are consistent with existing studies which place bounds on  $m_{\tilde{g}}$ and $m_{{\tilde{\chi}}_1^0}$ using Tevatron data on jets $+ \met$~\cite{Alwall:2008va}. 

\subsection{Pre-selection cuts and backgrounds}
To begin we define a set of pre-selection cuts which roughly characterize the kinematic features of the signal.  This will allow us to focus on the most relevant backgrounds.  Since 
we study benchmark models with heavy squarks at $m_{\tilde{q}}=4~{\rm TeV}$ and $5~{\rm TeV}$, which then decay into an ordinary QCD jet  and a gluino jet, our signal is characterized by jets and missing energy.  We therefore require:
\begin{itemize}
\item $p_T (j_1) >$ 1.5 TeV 
\item $p_T (j_2) >$ 400 GeV
\item{$\met >$ 500 GeV.}
\end{itemize}
where we use $j_i$ to denote the $i$-th hardest jet.  With these pre-selection cuts there are three major sources of background:
\begin{enumerate}
\item The most obvious background  given the pre-selection cuts above is $Z/W+{\rm jets}$ where $Z\rightarrow \nu \bar \nu$
or $W\rightarrow l \nu$\footnote{While $W\rightarrow l \nu$ backgrounds can normally be removed from ${\rm jets}+\met$ samples via a cut on isolated
leptons, here we will find the lepton from the $W$ decay is often collimated with a jet and thus non-isolated.  While identifying collimated leptons may be possible (see, for example, \Refs{citeulike:5845297,Lillie:2007yh}), to be conservative we will simply consider leptons as part of jets when they are close enough to be clustered with them.}.  
It is worthwhile to emphasize that the  effects of this background are somewhat non-standard for heavy-squark signals.  
Normally the $Z/W+{\rm jets}$  background can be significantly reduced via a cut on how dijet-like an event is (e.g. the $\alpha$ variable of \Ref{Randall:2008rw}) as 
SUSY events, which usually have two or more sources of $\met$, look less dijet-like than SM events\footnote{It is also worth mentioning
that the cross section for $Z+{\rm jets}$  can be surprisingly large when the $Z$ is allowed to be collinear with a jet~\cite{Rubin:2010xp}.  For configurations like
these the events are, morally speaking, dijet QCD events where a jet radiates a $Z$, leading to a $\ln^2 p_T/m_Z$  enhancement.}.  Here however,
 because the BSM events we study appear dijet-like, such a cut cannot be imposed without removing a significant portion of the signal.  
 
\item Another important background comes from boosted  $t\bar{t}$ production. Here $\met$ can arrise when one top decays semi-leptonically and the resulting lepton becomes difficult
to distinguish from the $b$-jet (due to their collimation). Just as with the leptonic decays of the $W$ in $W+{\rm jets}$ production, here the lepton can become collimated with a jet.  As discussed in \Ref{Rehermann:2010vq},  the lepton and the $b$ quark from a boosted top decay will be within $\Delta R = 0.4$ of each other roughly 50$\%$ of the time when $p_T \sim$ 1 TeV, and nearly $100\%$ of the time when $p_T \sim$ 2 TeV.  Due to this collimation, the resulting leptons will not pass isolation cuts and the events could be resolved as simply jets + $\met$.  As with the $W+{\rm jets}$ background, here we will take a conservative approach and cluster non-isolated leptons into jets.

\item Finally, QCD multijet events can contribute to our background in two ways.   Dijet events can contribute real missing energy when they fragment into $b$-hadrons, which can decay
semi-leptonically.  Furthermore, dijet events can yield fake missing energy when they are mismeasured.  Here we will simulate the effects of the first type of contribution (semi-leptonic
$b$-hadron decays) and provide efficiency estimates for methods to reduce the second type of decay\footnote{ In the kinematic region we are considering, where $\met$ is large compared to jet $p_T$s, any mismeasurement
error is non-gaussian and would require a more detailed detector simulation than we can confidently provide.  Therefore, we do not attempt to 
simulate this sort of mismeasured event - instead, later we only provide efficiency estimates.}.
\end{enumerate}

Before proceeding, we note that in what follows our signal events are generated at parton level using \texttt{Madgraph v4.4.49}~\cite{Alwall:2007st}, showered in \texttt{Pythia v6.422}~\cite{Sjostrand:2006za}, and matched using the MLM procedure~\cite{Hoche:2006ph}.
Some of our backgrounds (multi-jets and $Z$ + jets) are simulated in \texttt{Sherpa 1.2.3}~\cite{Gleisberg:2008ta} using the package's automated CKKW matching~\cite{Catani:2001cc}, while
others ($t\bar{t}$ and $W$ + jets) are simulated using the same Madgraph/Pythia flow with MLM matching that we used to generate our signal.
When we have run checks comparing Madgraph with MLM matching to Sherpa with CKKW, we find they agree with each other within roughly $20\%$ for most distributions.  After generation, all events are clustered into jets by \texttt{Fastjet v2.4.2}~\cite{Cacciari:Fastjet,Cacciari:2005hq} using the anti-$k_T$ algorithm~\cite{Cacciari:2008gp} with $R=$ 0.7.

%After the preliminary cuts, the cross section is 0.22 fb (Sherpa) and 0.18 fb (Madgraph).

%We use \textbf{Pythia 6.4.21} to generate the dijet sample and find that after applying the preliminary cuts, the QCD background will be below 0.01 fb and completely negligible. Among the four cuts, the first two would reduce the QCD cross section to about 0.8 fb while the last two cuts either reduce the background by roughly a factor of 10. 

\subsection{$\met$-jet alignment cuts}
The initial signal  cross sections (after accounting for the gluino branching ratio), as well as the signal and background rates resulting from the pre-selection cuts introduced earlier, are shown in \Tab{table: efficiency}.  Using the pre-selection cuts we see that the signal rates are reduced by roughly $60\%$, but the background are brought to the fb level.  
Furthermore, as can be seen in \Fig{fig:met}, the signal and background are both primarily dijet like (since $\Delta\phi(j_1,j_2)\sim\pi$) and $\met$ is mostly aligned with the second hardest jet.

As discussed earlier, in addition to the cuts already imposed it is customary to apply a cut on jet-$\met$ alignment requiring $\Delta\phi(j,\met)\gtrsim 0.3$.  However, as can 
be seen on the right hand side of \Fig{fig:met}, and in \Tab {table: efficiency}, such a cut would reduce signal rates by $60-70\%$.  While the $\Delta\phi(j,\met)$ cut does reduce background (especially the QCD background, which goes down by $\sim96\%$), such a cut is impractical and we must relax it, replacing it with something else.
\begin{table}
\begin{center}
\begin{tabular}{|c|c|c|c|c|c|c|c|}
\hline
& Model 1& Model 2& Model 3& $Z+J$ & $W+J$ & $t\bar t$& QCD\\
\hline
No cuts & 4.26 (0.51) & 3.78 (0.45) & 1.78 (0.23) & - & - & - & -\\ 
\hline
Pre-selection (PS) & 1.72 (0.27) & 1.46 (0.23)& 0.78 (0.14) & 0.43 & 1.05 & 0.41 & 0.82   \\
\hline
PS \& $\Delta \phi(\met,j)>0.3$ & 0.67 (0.09)  & 0.47 (0.06)& 0.31 (0.05)& 0.24 & 0.54 &  0.01 & 0.03\\
\hline
PS \& $y_{1\rightarrow 2} > 2\cdot 10^{-3}$ & 1.13 (0.18) &  1.15 (0.18) & 0.73 (0.12) & 0.07 & 0.32 & 0.18 & 0.17 \\
\hline
PS \& $y_{1\rightarrow 2} > 2\cdot 10^{-3}$ & 0.97 (0.16) &  1.01 (0.16) & 0.68 (0.11) & 0.04 & 0.18 & 0.06 & 0.09 \\
\& $p_T(j_3) > 100~{\rm GeV}$& & & & & & &\\
\hline
\end{tabular}
\caption{Signal (for $m_{\tilde{q}}=4(5)~{\rm TeV}$) and background cross sections, in fb, in the presence of various cuts.
\label{table: efficiency}
}
\end{center}
\end{table}

\FIGURE{
\begin{tabular}{ccc}
\includegraphics[scale=0.4]{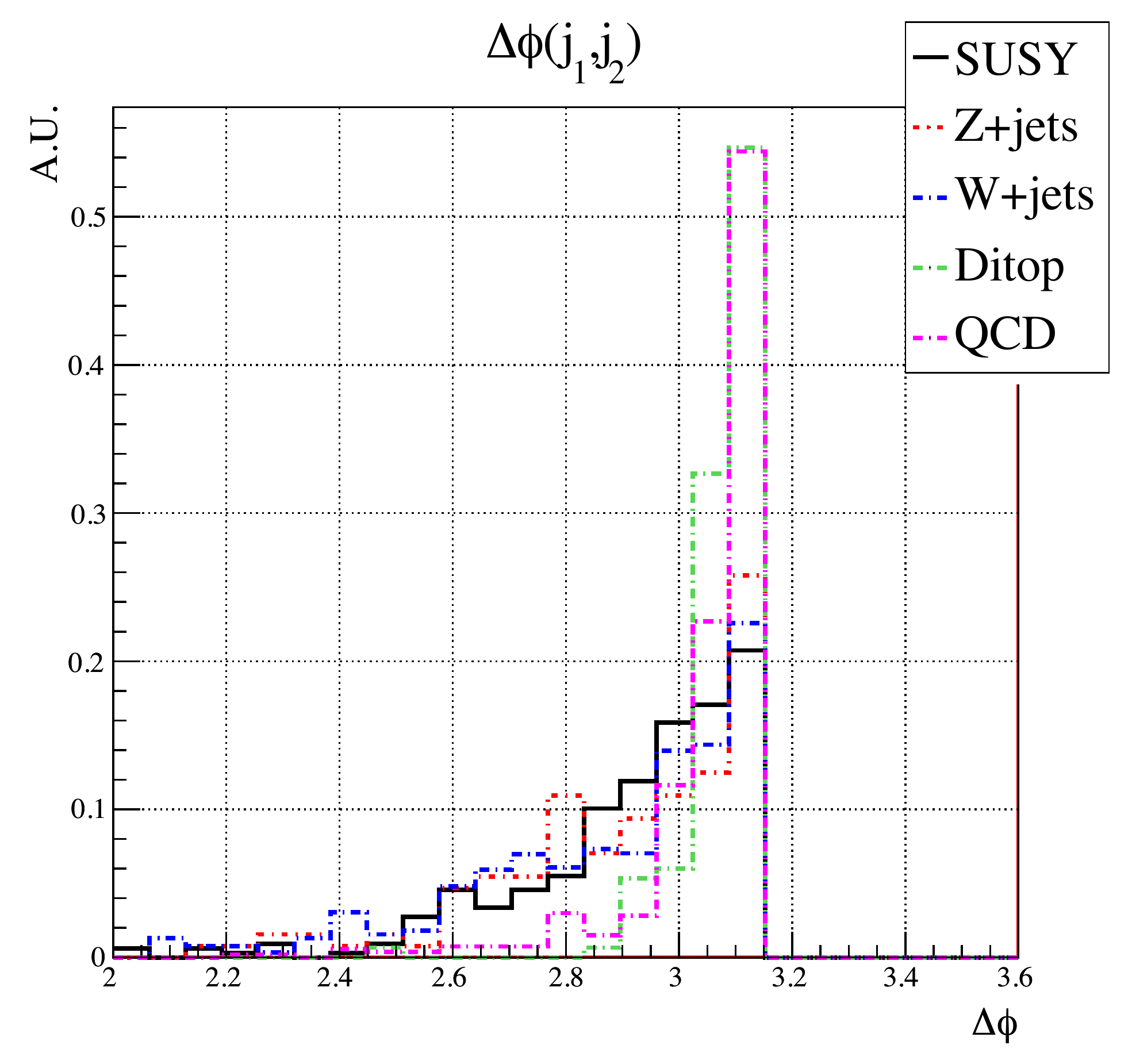}& 
\includegraphics[scale=0.4]{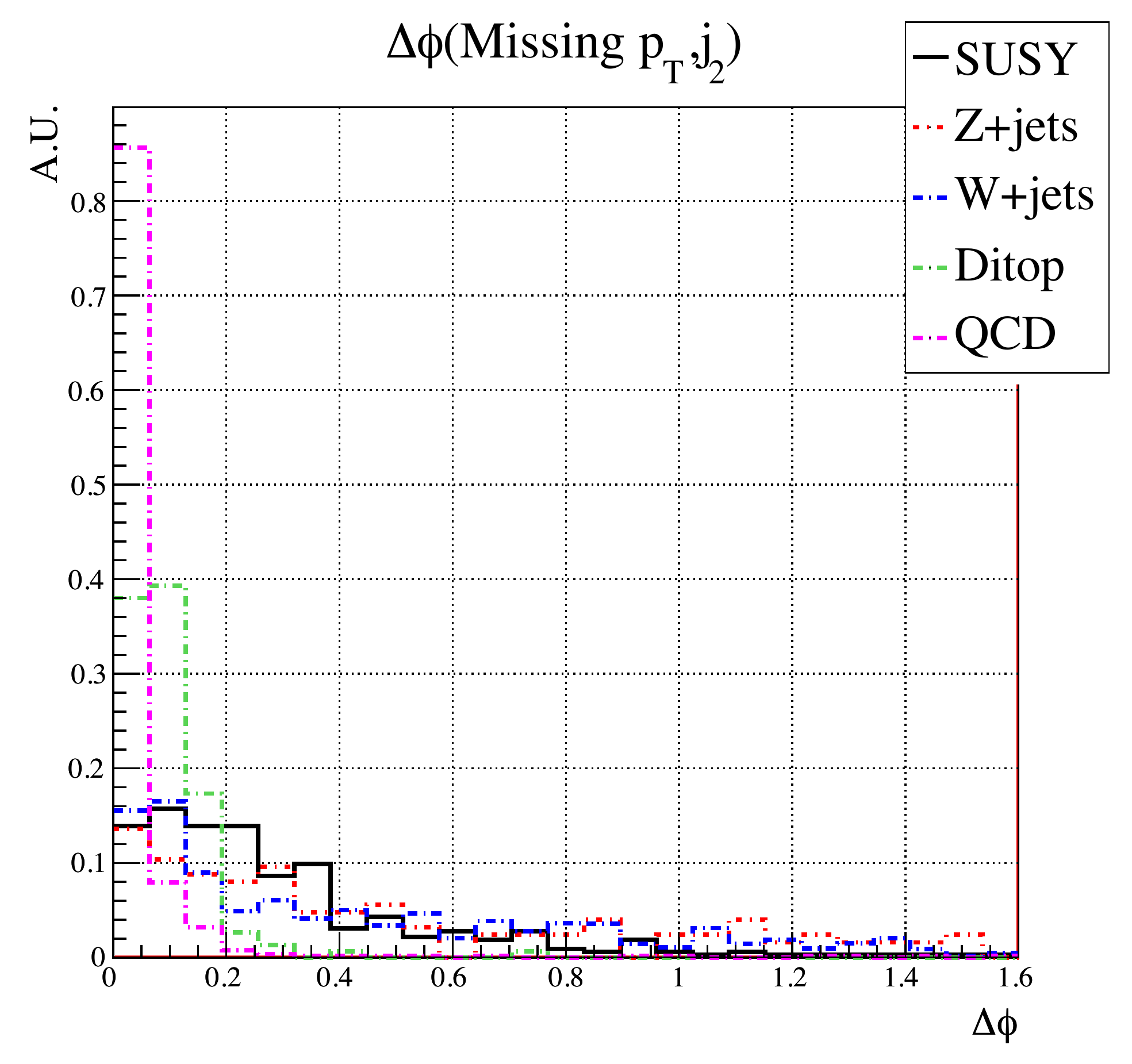} &
\end{tabular}
\caption{$\Delta\phi(j_1, j_2)$ and $\Delta\phi(\met, j_2)$ distributions on the left and right after the pre-selection cuts have been applied.  The distribution 
labeled ``SUSY'' is for a $m_{\tilde{q}}=4~{\rm TeV}$ squark decaying to light flavor: $\tilde{q}\rightarrow q\bar q\chi_0^1$ (Model 1).  However, we note that the 
distributions using other gluino decay modes are quite similar.  Note that all histograms are normalized to the same area.}
\label{fig:met}
}

%\FIGURE{
%\begin{tabular}{ccc}
% \includegraphics[scale=0.4]{ptj1.pdf}&
%\includegraphics[scale=0.4]{ptj2.pdf}&\\
%\includegraphics[scale=0.4]{ptj3.pdf}&
%\includegraphics[scale=0.4]{ptj4.pdf}& \\
%\end{tabular}
%\caption{The pt distributions of the four hardest jets. We normalize all the distributions to have one event.}
%\label{fig: pt}
%}

\subsection{Jet substructure variables}
As emphasized in the introduction, the gluino jets in our signal events have a rich substructure. Here we will investigate observables
sensitive to this substructure and use them to replace cuts on $\Delta\phi(\met, j)$.
 
%When energy depositions in a detector are mismeasured an ordinary QCD multijet event can become imbalanced
%and appear to contain $\displaystyle{\not} E_T$.  This is rare, especially for larger values of $\displaystyle{\not} E_T$, but because
%the cross section for QCD multijet events is so large, jet mismeasurement can be an important  background for BSM physics.
%
%Fortunately, these events share a distinguishing characteristic -- when a jet is mismeasured $\displaystyle{\not} E_T$ will 
%be aligned with the jet's direction, and by vetoing this sort of event a large fraction of the mismeasurement background can
%be removed.  However, as we saw in the preceding section, as long as $m_{\tilde{g}}\ll m_{\tilde{q}}$ the gluino from the squark decay will be boosted, forcing
% $\displaystyle{\not} E_T$ to be aligned with a jet.  Therefore, since a cut on $\displaystyle{\not} E_T$/jet alignment 
%would remove a large portion of the signal  we must find another tool helpful in distinguishing boosted gluinos from 
% events with mismeasured objects. 

The basic idea behind jet substructure  methods is that the 
distribution of constituent cells/particles
inside the jets of boosted heavy objects (like the gluino) 
is different than in ordinary QCD jets.  Basically, boosted objects tend to undergo a decay at a fixed order to partons of roughly equal energy (e.g.the top decay $t\rightarrow 3j$), while the jets from light particles radiate in a probabilistic fashion and with a strong energy hierachy.   This observation has been used  to identify and study
boosted EW gauge bosons~\cite{Seymour:1993mx,Butterworth:2002tt,Butterworth:2007ke,Cui:2010km,Han:2009em},  higgses~\cite{Butterworth:2008iy,Kribs:2009yh,Hackstein:2010wk,Kribs:2010hp,Plehn:2009rk,Kribs:2010ii}, tops~\cite{Kaplan:2008ie,Thaler:2008ju,Almeida:2008yp,Almeida:2008tp,Ellis:2009su,Ellis:2009me,Almeida:2010pa,Krohn:2009wm,Plehn:2010st}, and other 
exotica~\cite{Butterworth:2009qa,Falkowski:2010hi,Chen:2010wk,Katz:2010mr,Englert:2010ud}.
For a review, see \Refs{Abdesselam:2010pt,Salam:2009jx}.

A particularly simple observable sensitive to jet substructure is the $z$ variable of \Ref{Thaler:2008ju}.  To calculate  $z$ for a jet, 
one takes its constituents, reclusters them using the $k_T$-algorithm~\cite{Catani:1993hr,Ellis:1993tq}, and unwinds the clustering one step 
so that there are two subjets, $A$ and $B$.  Then, $z$ is defined as
\beqs
z=\frac{\min(E_A,E_B)}{E_A+E_B}
\eeqs
where $E_A$ and $E_B$ are the energies of the two subjets.  When the $k_T$-algorithm acts on the constituent  four-momenta (e.g. calorimeter cells)
of a jet, it does so by computing a distance between each pair of four-momenta using the metric
\beqs
\label{eq:d12}
d_{ij}=\min(p_{Ti}^{-2},p_{Tj}^{-2})\left(\frac{\Delta R}{R_0}\right)^2,
\eeqs
for $\Delta R$ the angular distance between two jets and $R_0$ a constant.
The smallest of these distance measures is chosen, and the four-momenta associated with it are combined together.  In this
way a jet is built up in stages, from soft to hard and in angle from near to far.
Now,  the angle and energy sharing of the radiation emitted by quarks and gluons has a soft/collinear singularity\footnote{i.e. a gluon tends to split into two gluons, where one is
very soft and/or collinear with the other.} and thus $z$, which measures this energy sharing is  $\sim 0$ for ordinary QCD jets, 
while boosted heavy objects, which have no such singularity~\footnote{e.g. because $h\rightarrow b\bar b$ decays 
isotropically in its rest frame it has no singularity when one of the $b$s becomes soft.}, yield $z\sim 1/2$.

The distributions of $z$ for our signal and background processes are shown in \Fig {fig:zdist}. 
Here we see that, as expected, 
the hard splitting present in the signal processes can be distinguished from the soft/collinear splitting of QCD. 
%Furthermore, 
%we note that the processes that have the most substructure (i.e. where the number of light partons in the leading order decay of the gluino is largest)
%tend to be more distinct than those with less.
\FIGURE{
\includegraphics[scale=.35]{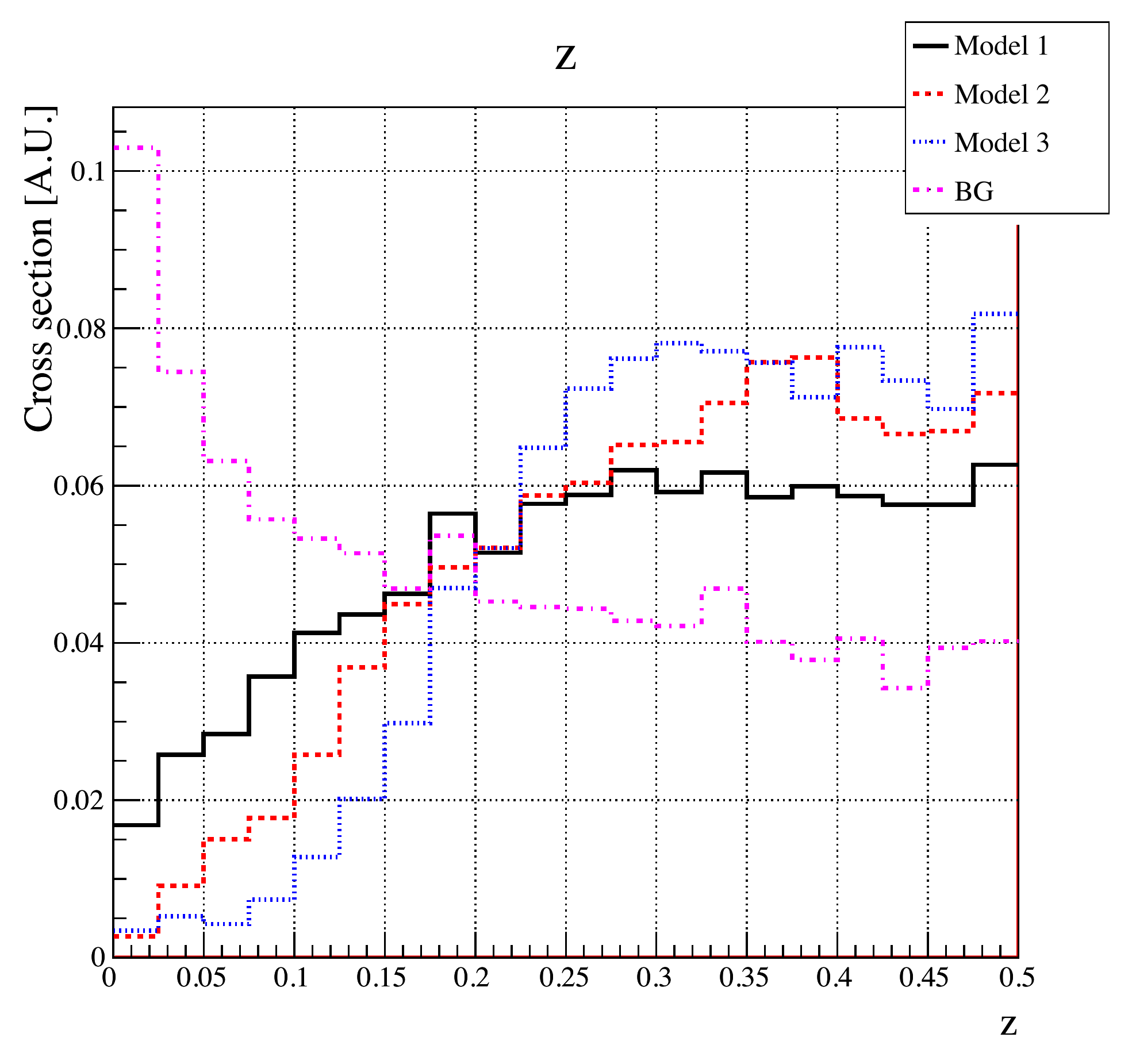} 
\includegraphics[scale=.35]{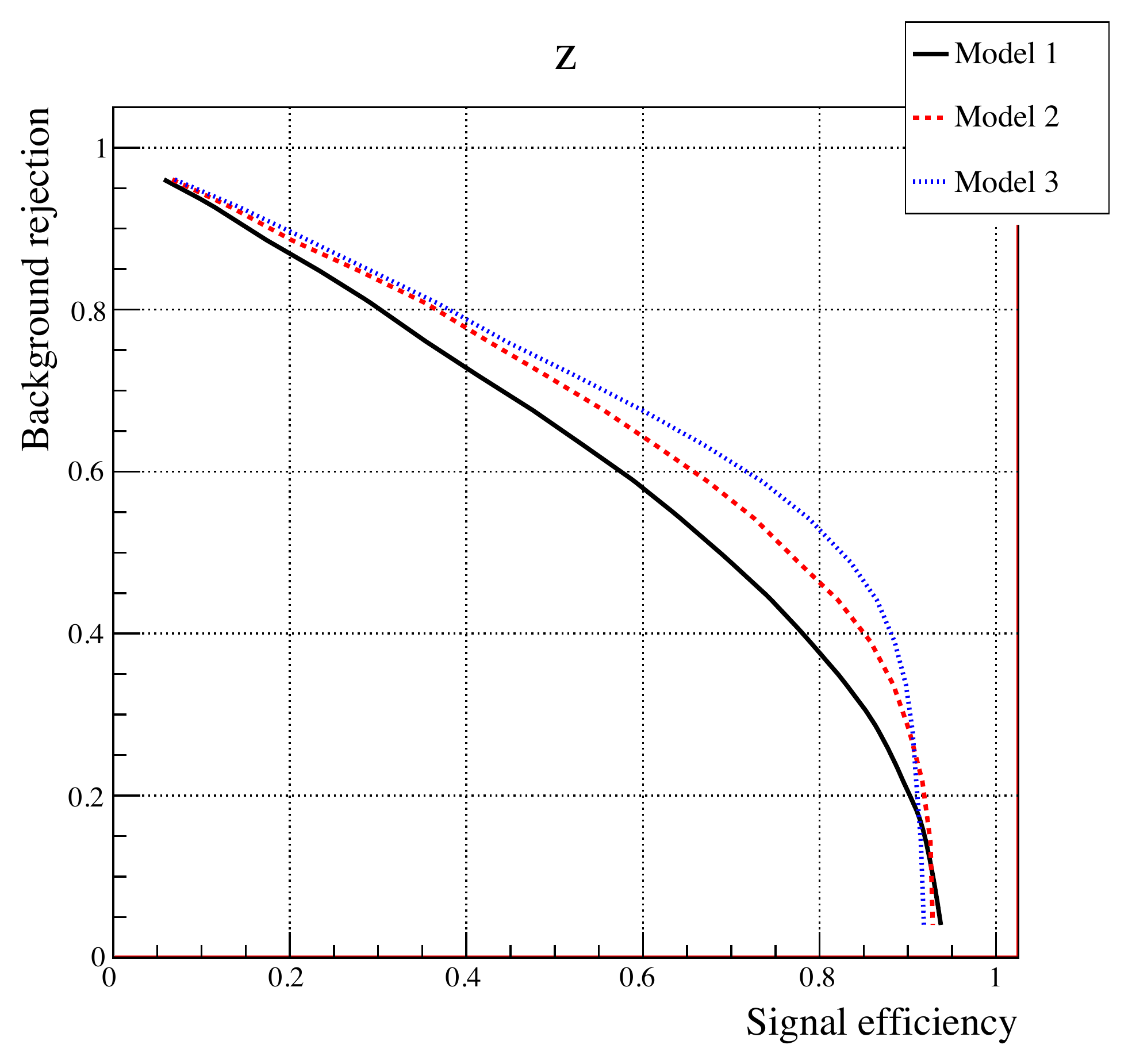} 
\caption{On the left, the distribution of the $z$ variable of \Ref{Thaler:2008ju}, and on the right, the signal and background 
efficiencies obtained with it.\label{fig:zdist}}
}

However, it is possible to do better.  The $z$ variable measures only the energy sharing of the subjets when a jet is broken in two
- it only provides minimal information about the angular dependence\footnote{The $k_T$ algorithm makes use of angular
information when constructing the two subjets.}.  However, the $y$ variable introduced in \Ref{Butterworth:2007ke}
\beqs
\label{eq:ysplit}
y_{1\rightarrow 2}=\frac{d_{1\rightarrow 2}}{p_T^2}
\eeqs
for $d_{1\rightarrow 2}$ the scale at which the initial jet is broken into two (see Eq.~\ref{eq:d12}) contains this information.
\FIGURE{
\includegraphics[scale=.35]{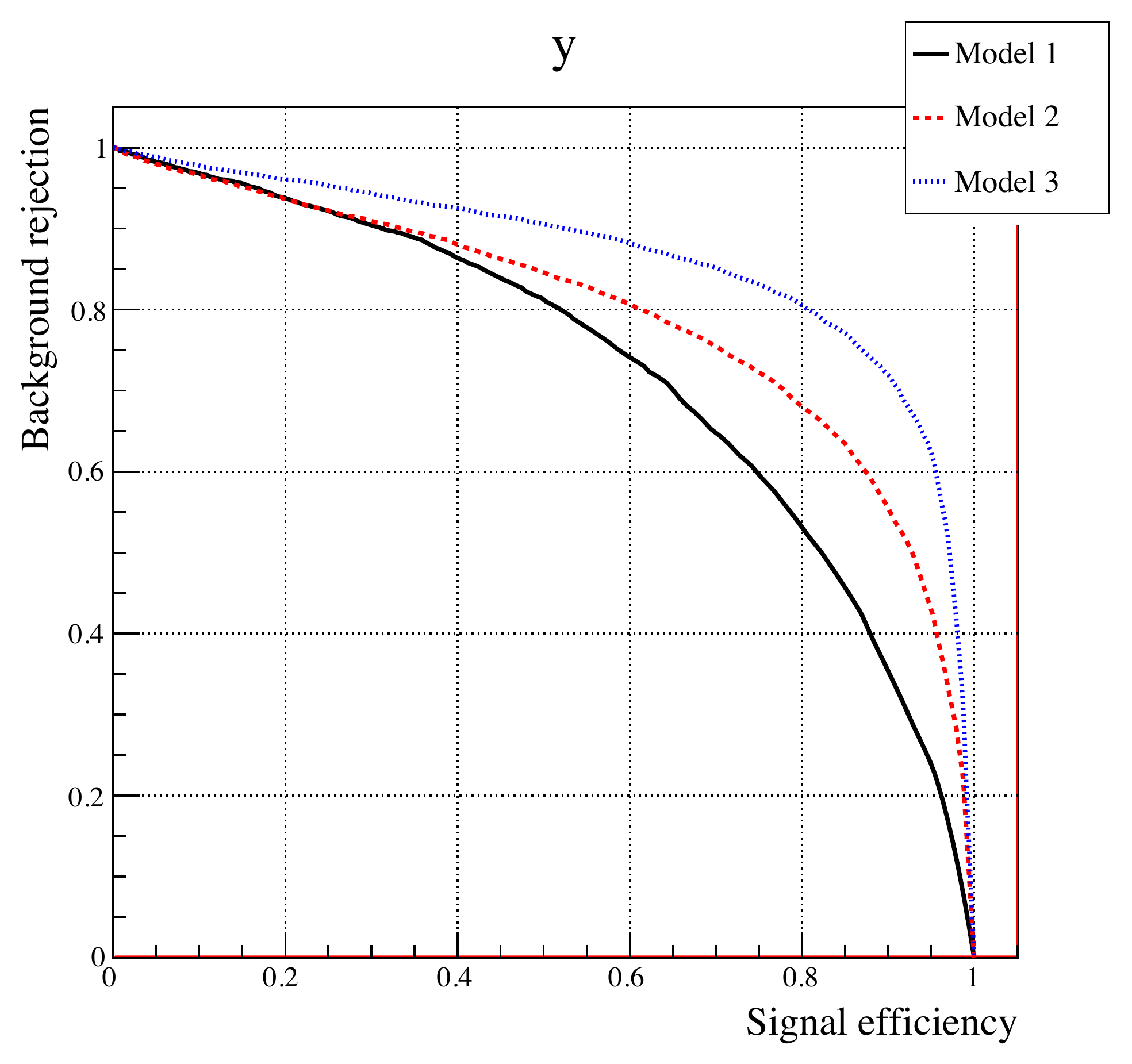} 
\includegraphics[scale=.35]{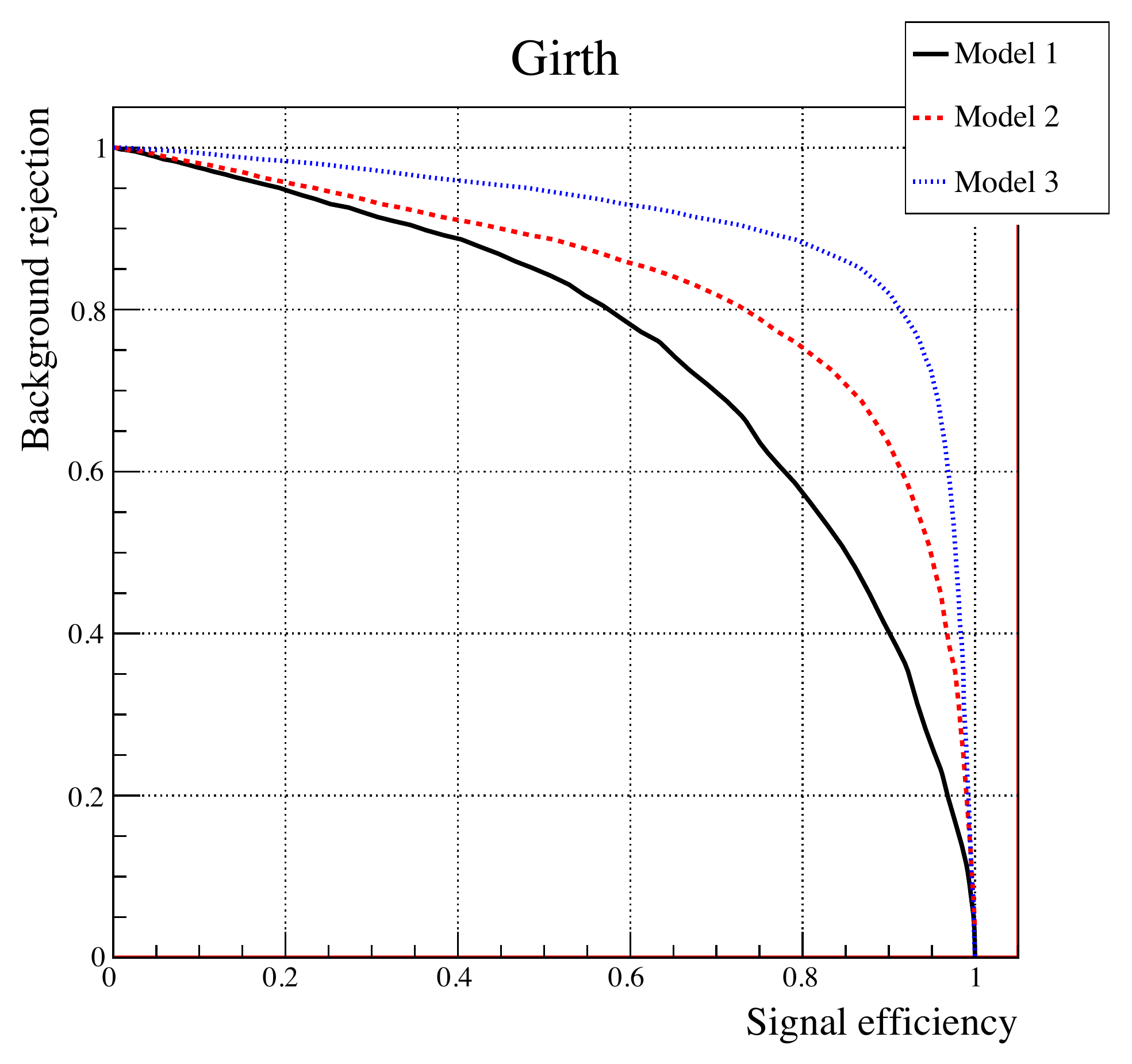} 
\caption{Signal and background efficiencies computed using, on the left, the of \Eq {eq:ysplit}, and on the right the girth
variable of \Eq {eq:girth}. \label{fig:girthdist}}
}
The resulting efficiency curves obtained from cuts using this variable are presented in \Fig {fig:girthdist}.  One can see a marked improvement 
over the results obtained with the $z$ variable.  Using $y_{1\rightarrow 2}$ it is possible to obtain over $90\%$ background rejection
with greater than $\sim40\%$ signal acceptance (and even $60\%$ signal acceptance for $\tilde{g}\rightarrow t\bar t$).

Finally, we note that it is possible to detect jet substructure without directly constructing subjets.  \Ref{Black:2010dq} introduced  a jet shape 
variable termed girth:
\beqs
\label{eq:girth}
g=\frac{1}{p_{T}}\sum r_i p_{T,i},
\eeqs
where the sum is taken over all jet constituents and $r_i$ is the distance from each to the jet center.  By making the replacements $E\rightarrow p_T$ and $\theta\rightarrow r$, one can see girth  is analogous to the `jet broadening'~\cite{Rakow198163} shape used at $e^+e^-$ colliders. 
The efficiencies obtained through the application of girth are shown in  \Fig{fig:girthdist}.  They are even better than the efficiencies obtained with the $y$ variable (although in 
what follows we will make use of the $y$ variable, as it is more widely used).
Before closing, we note that while these are not completely independent variables, they have different sensitivities and sensitivities to contamination, and thus
a more detailed experimental study is needed to optimize a cut on substructure. In closing, we note that the recently introduced $N$-subjettiness~\cite{Thaler:2010tr,Kim:2010uj}
variables might be also be useful for this purpose.

\subsection{Final cuts}
We now apply a substructure cut ($y_{1\rightarrow 2}> 2\cdot 10^{-3}$) in addition to the pre-selection cuts used earlier to find the efficiencies listed in \Tab{table: efficiency}.
From these efficiencies we see that the substructure cut removes a significant amount of the background (the QCD rate drops $79\%$) while much of the signal is retained (only dropping $20-30\%$).   However, the background levels are still worrisome, in particular for $W+{\rm jets}$.

Fortunately, we can still cut on the subleading jets.  Even though our signal is dominantly dijet like, the gluino produced in association 
with the squark decays largely into jets which are harder than those characteristic of the background.  This can be seen in \Fig{fig:thj},
where we show the $p_T$ of the third hardest jet in the events.  After imposing a cut on the third hardest jet (we use $p_T(j_3) > 100~{\rm GeV}$) 
we find the background has been reduced to the point where all three $m_{\tilde{q}}=4~{\rm TeV}$ models can be seen at $5\sigma$ in $10~{\rm fb}^{-1}$ of data
(see \Tab{table: efficiency} and \Tab{tab:finalsb}).  It is interesting to note that that after all of these cuts the three backgrounds contribute comparably.

Finally, we observe that while the cuts so far are insufficient to see $m_{\tilde{q}}=5~{\rm TeV}$ squarks in $100~{\rm fb}^{-1}$ of data, 
this can be remedied by increasing the jet $p_T$ cuts slightly to account for the higher squark mass.  Indeed, we find that simply increasing the 
cut on $p_T(j_1)$ from $1.5$ to $2~{\rm TeV}$ we reduce the background enough (by $86\%$) to see the heavier squarks\footnote{We emphasize though that
this higher cut on the leading jet $p_T$ is detrimental to the search for $m_{\tilde{q}}=4~{\rm TeV}$ squarks, and for these a cut of $1.5~{\rm TeV}$ should be used.}.  The final cross sections and LHC sensitivity can be see in \Tab{tab:finalsb}.
\FIGURE{
%\begin{tabular}{ccc}
\includegraphics[scale=0.5]{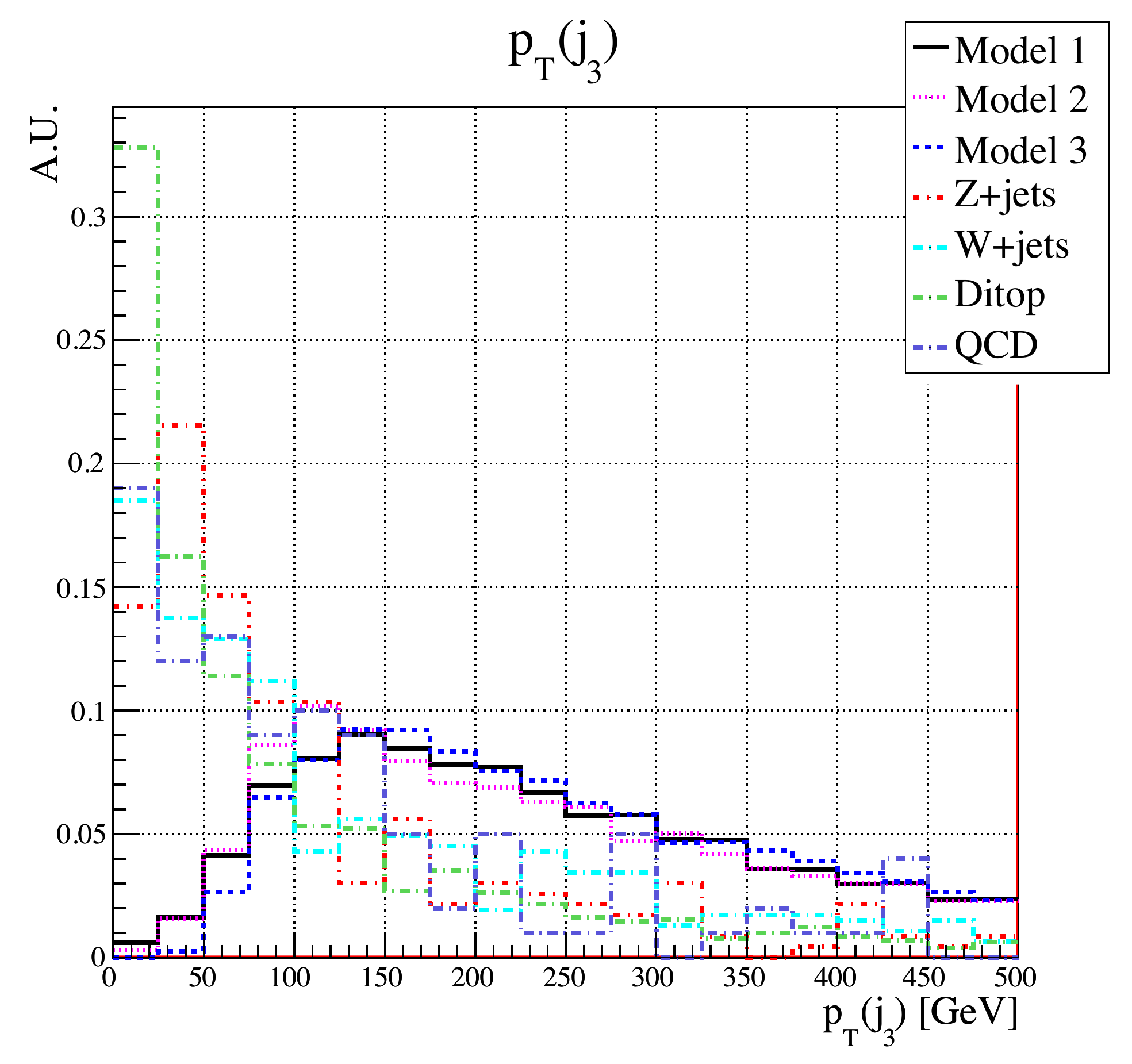}
%\end{tabular}
\caption{$p_T$ of the third hardest jet for different signal models (all with $m_{\tilde{q}}=4~{\rm TeV}$) after applying the pre-selection
cuts defined earlier, along with a cut requiring $y_{1\rightarrow 2}>2\cdot 10^{-3}$ for the second hardest jet.}
\label{fig:thj}
}

\begin{table}
\begin{center}
\begin{tabular}{|c|c|c|c|c|c|c|c|}
\hline
& Model 1& Model 2& Model 3\\
\hline
$\sigma_{\rm S}$& 0.97 (0.11) & 1.01 (0.11) & 0.68 (0.08)\\
\hline
$S/\sqrt{B}$& 5.0 (4.9) & 5.3 (4.9) & 3.5 (3.6)\\
\hline
\end{tabular}
\caption{Signal cross section and significance for $m_{\tilde{q}}=4(5)~{\rm TeV}$ at ${\cal L}=10(100)~{\rm fb}^{-1}$.  To arrive at these numbers
we use the preselection cuts defined earlier (increasing the cut on $p_T(j_1)$ to $2~{\rm TeV}$ for $m_{\tilde{q}}=5~{\rm TeV}$) and 
required $p_T(j_3)>100~{\rm GeV}$ and $y_{1\rightarrow 2}(j_2)>2\cdot 10^{-3}$.
Note that, after applying these cuts we found background cross sections of $\sigma_{\rm B}=0.37 (0.05)~{\rm fb}$.\label{tab:finalsb}
}
\end{center}
\end{table}

%
%\FIGURE[h]{
%\begin{tabular}{ccc}
% \includegraphics[scale=0.4]{m12res.pdf}&
%\includegraphics[scale=0.4]{m34res.pdf}&
%\end{tabular}
%\caption{Left: the invariant mass distributions of the first and second hardest jets; Right: the invariant mass distributions of the third and fourth hardest jets. Both distributions are after the preliminary cuts and the jet mass cut $M(j2) > 100$ GeV. }
%\label{fig: inv mass}
%}

\section{Conclusion}
\label{sec: conclusion}
Here we have considered a relatively unstudied region of SUSY parameter space exhibiting interesting collider phenomenology.  In the region we investigated first 
generation squarks are very heavy, leading to signal events which appear dijet like with $\met$ aligned with a jet.  We found that while we were able to go far in reducing the background to this channel via  cuts on jet $p_T$s and $\met$, much of our signal was removed by jet/$\met$ anti-alignment cuts used by LHC experiments to guard against jet mismeasurement errors.

We suggested replacing these cuts by a measurement  of jet substructure.  This allowed us to still remove much of the background while retaining a substantial portion of the signal.  After applying this and the aforementioned cuts, we demonstrated heavy squarks could be seen at $m_{\tilde{q}}=4(5)~{\rm TeV}$ in ${\cal L}\sim 10(100)~{\rm fb}^{-1}$ of data.  Interestingly, in the end all of our background contributed equally, motivating the need for a careful experimental study of the relevant efficiencies (so as to avoid an unfortunate Altarelli cocktail).

In closing, we note that it would be interesting to see if other signals could be enhanced by replacing anti-alignment and similar cuts on jet quality cuts by measurements of jet substructure.  In addition, 
the backgrounds we encountered could potentially be reduced even more through cuts on non-isolated leptons.  We hope this serves as further motivation for experimental study of this interesting configuration.

\acknowledgments{The authors would like to thank N. Arkani-Hamed for suggesting this topic, and G. Salam, M. Schwartz, B. Tweedie,  and Z. Han for useful discussions. J.F. is supported by the NSF under grant PHY-0756966. D.K. is supported by a Simons postdoctoral fellowship and by an LHC-TI travel grant. A.T. is supported in part by the DOE under Grant No. DE-FG02-90ER40560. L.-T.W. is supported by the NSF under grant PHY-0756966 and the  DOE Early Career Award under grant DE-SC0003930.  }

\bibliography{hs}
\bibliographystyle{jhep}
\end{document}